# A Lightweight and Scalable Physical Layer Attack Detection Mechanism for the Internet of Things (IoT) using Hybrid Security Schema

**Reza Fotohi**[1]  . **Hossein Pakdel**[2]

**Abstract** The Internet of Things, also known as the IoT, refers to the billions of devices around the world that are now connected to the Internet, collecting and sharing data. The amount of data collected through IoT sensors must be completely securely controlled. To protect the information collected by IoT sensors, a lightweight method called Discover the Flooding Attack-RPL (DFA-RPL) has been proposed. The proposed DFA-RPL method identifies intrusive nodes in several steps to exclude them from continuing routing operations. Thus, in the DFA-RPL method, it first builds a cluster and selects the most appropriate node as a cluster head in DODAG, then, due to the vulnerability of the RPL protocol to Flooding attacks, it uses an ant colony algorithm (ACO) using five steps to detect attacks. Use Flooding to prevent malicious activity on the IoT network. In other words, if it detects a node as malicious, it puts that node on the detention list and quarantines it for a certain period of time. The results obtained from the simulation show the superiority of the proposed method in terms of Packet Delivery Rate, Detection Rate, False Positive Rate, and False Negative Rate compared to IRAD and REATO methods.

## 1 Introduction

The Internet of Things has affected all aspects of our lives and developed over time due to the advancement of technologies such as wireless communications, sensors, and large-scale computing. There are many uses for the Internet of Things, ranging from common home appliances to large-scale and critical safety systems [1-5]. The number of IoT devices can be expected to exceed 50 billion by 2030. However, despite the many benefits of the Internet of Things, there are potential intruders who can compromise the normal behaviour of sensors. This necessitates safe and secure IoT design. One of these dangerous attacks is the flooding attack. A flooding attack falls into the category of DDoS attacks. This attack tries to minimize the capacity of the desired thing to respond to the request of other normal things by sending RREQ messages too much and keeping things busy and prevent other things from accessing this service. In this paper, to solve such problems, a method based on ACO and DODAG is used to prevent dangerous attacks such as flooding to have a safe routing in the network.

✉ Reza Fotohi*
R_fotohi@sbu.ac.ir; Fotohi.reza@gmail.com

✉ Hossein Pakdel
Hos.pakdel.eng@iauctb.ac.ir

[1] Faculty of Computer Science and Engineering, Shahid Beheshti University, G. C. Evin, Tehran 1983969411, Iran
[2] Department of Computer Engineering, Central Tehran Branch, Islamic Azad University, Tehran, Iran

The proposed DFA-RPL method consists of two main parts, in the first part, clustering and selecting the cluster head to create DODAG is done, and in the second part, Flooding attack detection is done using ACO algorithm. The first part is used to create an optimal DODAG that has the best nodes as a parent. The task of the second part is to detect malicious things, which are most likely Flooding nodes that try to disrupt network operations by sending a large number of messages. These malicious nodes are detected using the ACO algorithm and using five steps. Its destructive activity in the network is prevented by the sound that if the node is found to be malicious, that node will be placed on the detention list and will be quarantined for a certain period of time. The proposed DFA-RPL scheme can efficiently create an optimal DODAG and manage the Flooding attack on the network.

The organization of the different sections of this paper is as follows: Section 2 describes the latest related work related to the topic of this paper. The proposed DFA-RPL approach with all its details and subsections is given in Section 3. Section 4 of this paper presents the performance evaluation in which the simulation results of the proposed method and the other two methods were compared and discussed. Conclusion This article is presented in the final section.

## 2  Related work

The IoT, as an emerging technology, is expected to offer promising solutions for the transformation and realization of services provided by various applications [6-9]. This network can be defined as the combination of the physical world with the digital Internet and virtual world on a global scale. These networks face many challenges, especially in communications and networks due to their inherent limitations, fear of resource inefficiencies and limited capabilities. Remote needs assessment makes sensor networks and low power networks one of the most important IoT technologies.

Hatzivasilis et al. [10] proposed a trust management method and channel validation system that is a new system for secure routing in wireless ad-hoc networks. The purpose is to maintain network security and high system performance. SCOTRES criteria include energy efficiency criteria that prevent energy loss and increase node load, which increases grid life. It also introduces a new concept that protects network nodes and improves load balance, and finally the channel health standard, which estimates the condition of the channel between nodes and prevents blockage and congestion in some areas.

Jiang et al. [11] proposed a secure routing method aimed at identifying malicious nodes that increases the quality and reliability of communication between independent nodes through the routing protocol.

Lahbib et al. [12] designed a model of informed trust and reliable communication in the RPL protocol to prevent internal attacks. The purpose is to ensure trust among object entities and to provide quality assurance of service when building and maintaining a network routing topology.

Devi et al. [13] proposed a central content load-based routing method. This algorithm transmits data based on content. Routing related data in the same direction reduces traffic. Therefore, the latency is reduced, which reduces traffic and saves battery power.

Gawade et al. [14] identified gaps in the implementation of the RPL secure routing protocol that triggered security attacks and provided solutions to overcome this problem. This protocol improves reliability and security.

Hatzivasilis et al. [15] developed a trust-based system for secure routing and secure licensing. The system consists of three main components: 1- An encryption service that performs authentication, message integrity and confidentiality 2- A trust-based routing system that protects communications against ad-hoc routing attacks.0- Policy-based on the framework control that provides the permissions.

Glissa et al. [16] proposed the SRPL secure routing protocol. The main purpose of SRPL is to prevent incompatible nodes that have destructive changes to message control such as node validation and corrupt networks by creating a fake topology.

Xia et al. [17] developed a method called LX-AGV that prevents fraudulent attacks within the IoT network. Logic is based on a well-known trust mechanism that gradually identifies and manages malicious nodes as well as defends against internal fraudulent attacks. The goal of the existing routing protocol is successful in terms of transmission speed, and the average latency and loss ratio work better. This mechanism has been proposed to defend against unpredictable attacks.

Djedjig et al. [18] proposed the MRTS secure routing protocol. In this paper, a new RPL scheme is presented that supports the definition of shared security. This method uses reliable joint evaluation between different nodes within the network. Node behaviours are determined by the components of selfishness, energy, and honesty. Based on this, it is possible to check the validity value of all nodes that are directly connected to the current node in the neighbourhood and choose the safest communication route with a high score.

To prevent network layer attacks and attacks on routing protocols, they proposed a deep learning-based approach that can detect high-distribution data based on large-scale data. The mechanism proposed in [19] can successfully detect a flooding attack and eliminate it from the operation cycle.

In [20], the solution designed to fit the NPS architecture is authenticated using a real-world test platform and built by an NPS prototype that receives open data in real time through a set of compatible sources. Slowly This method, called REATO, is used to detect and thwart a DoS attack compared to the IoT middleware known as NPS. Work has begun on the need to discover a solution to protect an IoT system from DoS attacks, taking into account all potential contingencies that may occur.

# 3 The proposed DFA-RPL schema

In this paper, we seek two goals:
- Creating a cluster in the RPL protocol
- Detecting the flooding attack and identifying the secure route

The proposed Discover the Flooding Attack-RPL (DFA-RPL) method first builds a cluster and selects the most suitable node as the cluster head in DODAG. Then, the Ant Colony Optimization (ACO) algorithm is used to detect these attacks due to the vulnerability of the RPL protocol in flooding attacks. In the proposed method, a safe path for data transmission is selected for greater assurance even upon the detection of the flooding attack. Figure 1 displays the overall architecture of the proposed method based on clustered RPL and DODAG, in which the flooding attack nodes are present. The green-colored nodes in the figure are the cluster head nodes, and the zero-level cluster head node is shown in purple. The main root is DODAG, which is considered a safe node. The blue nodes are the members of each cluster in DODAG, and the red nodes are the flooding attack malicious nodes. Based on the proposed DFA-RPL method, the flooding attacks are detected after the clustering and the selection of the best cluster head in DODAG.

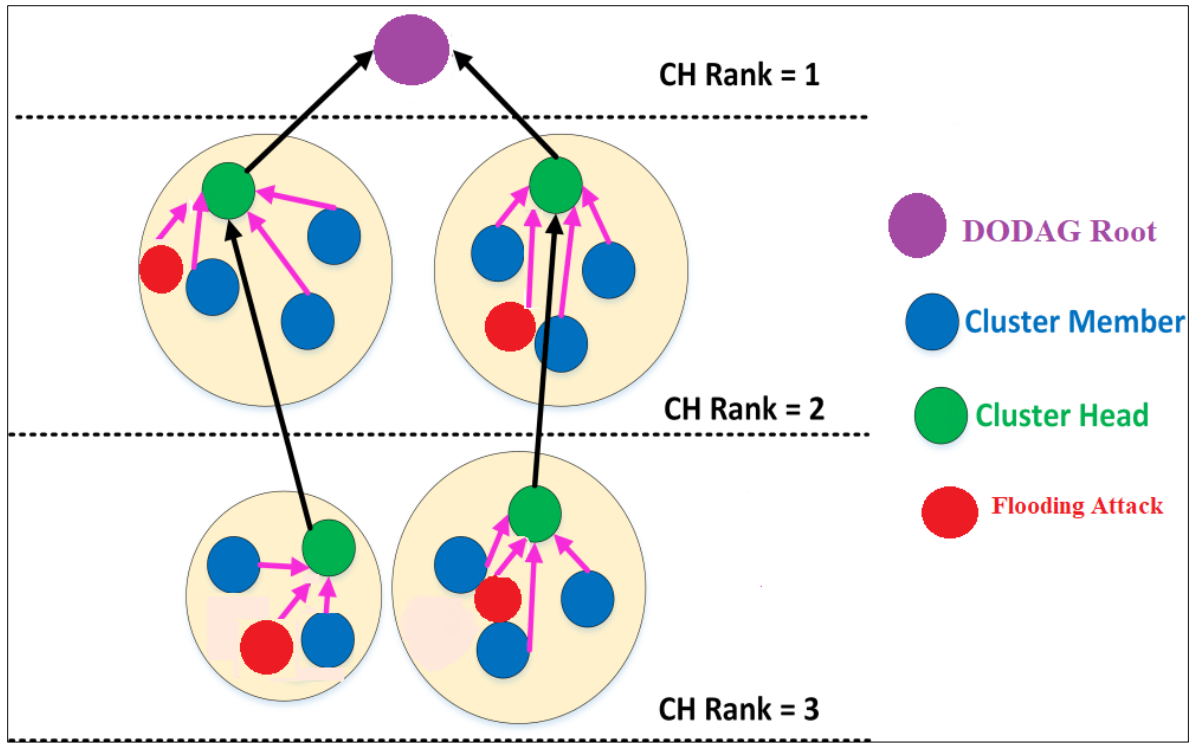

**Fig. 1**. Overall architecture of the proposed DFA-RPL method.

## 3.1 Assumptions of the proposed DFA-RPL method

In We have considered some parameters as default for our proposed method, as follows:
- The proposed IoT network contains N IoT nodes with $N = \{N_1, N_2 \ldots N_n\}$ and with limited sources.

- The things or nodes can be sensors or mobile things. These things are within the network range and have homogeneous sources.
- All the things are equipped with GPS and are aware of their own position.
- By default, there is no central controller in the IoT network.
- Communication between the nodes is performed by the RPL routing protocol, and RPL creates a virtual DODAG on the network topology.

## 3.2 Clustering and selecting the cluster head in DFA-RPL

The DFA-RPL method begins with the process of selecting the best node as the cluster head. The cluster head node is selected as the most central node using three criteria: number of hops, energy, and distance criterion. The number of hops refers to the number of jumps required for sending data from the cluster members to the cluster head. The smaller this criterion, the faster the data transmission. The energy criterion is important because the cluster head is responsible for collecting data from all the cluster members and sending them to the DODAG main root. The higher the energy of the cluster head, the more the data it can transmit, and the less the energy lost during data transmission.

In the first step, all the things send a "hello" message containing information such as their residual energy and position to the DODAG root. Based on all the member information, the DODAG root selects a number of high-energy nodes, which higher energy relative to the other things, in each area, and their distance and Received Signal Strength Indicator (RSSI) with respect to the neighboring nodes are calculated. The residual energy ($E_{RE}$) is calculated using Eq. (1).

$$E_{Co} - E_{Pr} = E_{RE} \tag{1}$$

In Eq. (1), $E_{Pr}$ is the initial energy of the thing, and $E_{Co}$ denotes the energy consumed by the thing. The radio energy model is used to determine the energy consumption level. In this model, the energy associated with transmitting k bits of data between a transmitter thing and a receiver thing that are $d$ meters apart is calculated using Eq. (2):

$$\begin{aligned} E_{Co} &= ETX + E_{RX} \\ E_{TX} &= E_{elec*} + E_{amp*} \\ E_{RX} &= E_{elec*} \end{aligned} \tag{2}$$

In Eq. (2), $E_{TX}$ represents the energy consumed by the thing for transmitting data. The parameter $E_{RX}$ is equal to the energy consumed by a thing for receiving data. The energy required to operate the electronic components, i.e., the transmitter and the receiver is expressed by $E_{elec}$, and the energy required to amplify the transmitted signal is shown using $E_{amp}$. The value of $E_{amp}$ depends on the data transmission distance. To calculate the distance between the high-energy node and other nodes, first, the transmission equation is in the form of Eq. (3):

$$\frac{POW_R}{POW_T} = A_r A_T \left(\frac{\lambda}{4D\Pi}\right)^2 \tag{3}$$

$$\lambda = \frac{S}{f}$$

$POW_R$ denotes the power for receiving data, and $POW_T$ is the power for transmitting data. Moreover, $A_r$ is the antenna gain for receiving data, and $A_T$ represents the antenna gain for transmitting data, which are expressed in decibels. Also, λ is the wavelength, which can be calculated using the node speed (S) and the frequency (f). Finally, D is the distance and is calculated according to Eq. (4), and $\pi = 3.14$.

$$D = \sum_{v \in V, v' \neq V} dist(v, v') \tag{4}$$

$$Dist = \sqrt{(x_1 - x_2)^2 + (y_1 - y_2)^2}$$

In Eq. (4), $(x_1, y_1)$ is the geographical position of the thing v, and $(x_2, y_2)$ denotes the geographical position of the neighboring thing v'. From (3), the RSSI is calculated according to Eq. (5):

$$RSSI = Log \frac{POW_R}{POW_T} * 10 \tag{5}$$

Sum up the distances calculated for the selected high-energy things. If the distance of a node from its neighbor is long, more energy and power are required for data transmission; hence, more energy is consumed. Therefore, the things with the smallest distance to the neighboring things and the root and with higher *RSSI* are selected as the cluster head. The higher the *RSSI* of a node, the higher the power of the signal received by that thing. In each area, the things with larger RSSI and smaller distance with respect to the neighboring things are selected as the optimal cluster head $CH_{opt}$. In other words, things with the highest energy, nearest to other things, and with the highest *RSSI* are selected, as expressed by Eq. (6):

$$CH_{opt} = \left(\frac{Max[E_{RE}n_1, E_{RE}n_2, E_{RE}n_i], Max[RSSI_{n1}, RSSI_{n2}, ...RSSI_{ni}]}{Min[D_{n1}, D_{RE}n_2...D_{ni}]}\right) \tag{6}$$

| **Algorithm 1:** Selecting the cluster head in the DFA-RPL method |
|---|
| **Input:** Selecting *i* random things with the highest residual energy in each area |
| **Output:** Selecting the optimal cluster head (the process $CH_{opt}$) |
| 1. Calculate the distance and RSSI for the selected things. |
| 2. Calculate the sum of the distances of each thing from all the neighboring things. |
| 3. Select the things with the smallest distance from the neighbors and the highest RSSI as the cluster head of each area. |
| 4. The thing selected as the cluster head and the neighbors of every cluster head are informed by the root. |
| 5. Transmission of DIO message for creating the DODAG |

In the proposed DFA-RPL method, after the completion of the cluster head selection process, the process of DODAG creation in each cluster begins. The process is as follows: each thing that has been selected as a cluster head transmits the DIO message to its neighboring things and announces the initiation of DODAG creation. The DIO message is one that contains ICMPv6 packages, which are specifically used in the RPL protocol. All the neighboring things that have received the DIO message join the DODAG, and each thing predicts its own rank, based on which the next parent of each thing is selected. The thing selected as a parent transmits the DIO message again, and the DODAG is fully constructed.

In this technology, all the things consume energy. In the proposed method, if the life of none of the selected cluster heads ends and none of the cluster heads reaches the death state, each thing will select the cluster head of the subsequent step based on the rank value. In the DFA-RPL method, the rank and the residual energy are considered for determining the new cluster head. This minimizes the process of repeatedly selecting the cluster head node. The process of selecting the cluster head is performed only when a cluster head node has very low energy or reaches the death state. When a cluster head node is transferred to another cluster, a new cluster head is determined.

### 3.3 Flooding attack detection using ACO algorithms

In the proposed method, the ACO algorithm is used to detect flooding attackers. This operation is performed by the cluster head. Data such as IP, protocol type, destination address, and the number of requests received are kept in each thing. The ACO algorithm in the proposed DFA-RPL method is executed in 5 consecutive steps, as follows:

(1) evaluating each thing,

(2) fitness estimation,

(3) calculating the overall fitness,

(4) computing the maliciousness probability of the thing, and

(5) identifying the flooding malicious node.

These steps are performed in the cluster where the DODAG has been created. The workflow is as follows: the pheromone is calculated as in the regular ACO algorithm; in the proposed DFA-RPL method, the amount of pheromone is estimated in the first step to evaluate the things. The amount of pheromone is calculated using Eq. (7).

$$F = f_0 * \left(1 - \frac{\alpha * N_t}{N_{max}}\right) \quad (7)$$

In Eq. (7), $N_t$ is a variable representing the number of messages that have arrived at the cluster head, $N_{max}$ is a variable that indicates the maximum number of variables, and $\alpha$ is the tuning coefficient, which varies in the range $\epsilon[0,1]$. $f_0$ denotes the initial values of the parameter $F$. Since in the flooding attack, the intruder transmits a flood of packages, it obstructs the traffic in the network and, hence, causes the legal requests and messages of the users to be rejected. Therefore, in the proposed method, the number of messages is considered to be the most important criterion due to the nature of the flooding attack. After evaluating the amount of pheromone, the fitness estimate is calculated in the second step as in Eq. (8):

$$Fit_e = \left(\frac{1}{1+F}\right) \qquad (8)$$

The maliciousness probability of each thing is computed according to the fitness value, i.e., Eq. (7). Therefore, in the next step, the maliciousness probability of a thing as a flooding intruder is calculated using Eq. (8). In Eq. (8), T ($Fit_e$) is the total fitness value, which is obtained by adding up all the fitness values. The thing with a higher probability is considered a flooding intruder.

---

**Algorithm 2:** Detecting the flooding attack using the ACO algorithm

**Input:** $n$ things present in the network

**Output:** Detection of malicious things in the flooding attack

**Algorithm process:**

1. Population initialization:

2. Evaluation of things using (6)

3. Fitness estimation using (7)

4. Determination of the total fitness as $Tfit_e = fit_e(1), fit_e(2),\ldots$

5. Calculation of the maliciousness probability of each thing

6. Detection of the flooding malicious thing that has the highest probability

7. Broadcast of a warning message to the DODAG to introduce the malicious thing.

---

When the cluster head detects the presence of a malicious node, a warning message containing the ID of the flooding intruder thing is sent to the DODAG root so that it is broadcast to the things present in the network.

### 3.4 Adding the malicious thing to the detention list

In the proposed DFA-RPL method, when a cluster head detects a flooding attack malicious thing, it sends the thing's ID to the DODAG root so that it can be broadcast to all the things in the network. Then, corresponding cluster head and other cluster heads add the ID of the malicious thing to their detention list and reject all the messages received from that thing for a period of Θ.

### 3.5 Intruder reconsideration

Each object keeps the field containing the detention list for a malicious thing for a period Θ=4*RTT. RTT is the average time for a round of messages to be sent to and from the DODAG root. After the end of this time, the malicious thing is considered a valid node in one round under the supervision of the cluster head, and if it operates as a normal thing, it will be considered a valid thing. When a thing is considered a valid or normal thing, the DODAG root is informed so that the root and the other things in the network update the field related to this thing. If this thing or other things exhibit malicious behavior again, they are once again placed in the detention list, and all the things in the network make modifications accordingly. The proposed DFA-RPL method is a completely safe method for restricting a flooding attack in IoT. The steps in the DFA-RPL method are specified in the flowchart of Figure 2.

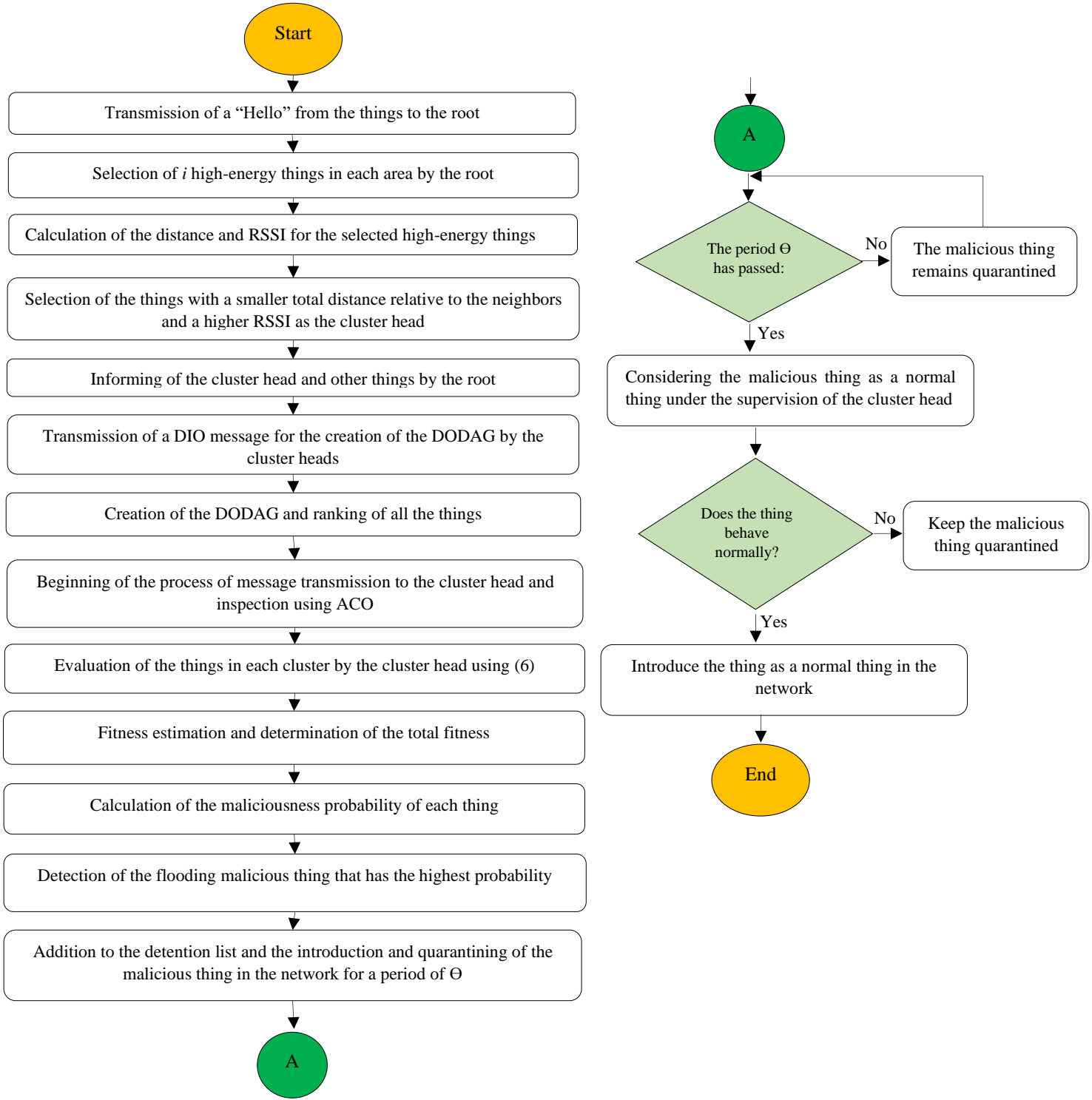

**Fig. 2**. Flowchart of the proposed DFA-RPL method.

# 4 Performance Evaluation

This section has two subsections called Performance metrics and Simulation results and analysis: The results extracted from the simulator are shown in the format of tables and graphs. It also compares the simulation results of the proposed DFA-RPL method with the last two methods that worked on the detection of malicious nodes under important criteria. The results were compared with three methods (IRAD [19] and REATO [20]). To demonstrate a feasibility study, the performance analysis of REATO, and IRAD has been divided into four parts: False Positive Rate, False Negative Rate, Detection Rate, and Packet Delivery Rate.

## 4.1 Performance Metrics

In this subsection, the concept and formula of all measure are explained in detail. Also, the meaning of the variables used in Eq. (9) is explained in Table 1.

**Packet Delivery Rate:** Total packets that were successfully received in the destination thing divided by the total packets that were successfully sent in the origin things. Multiply the result by 100 to get the percentage [21, 22]. This measure has been analyzed using Eq. (9).

$$APDR = \left( \left( \frac{1}{N} \right) * \left( \frac{\sum_{i=1}^{N} R_i}{\sum_{i=1}^{N} S_i} \right) * 100 \right) \quad (9)$$

**Table 1** Variables and their meanings

| Variables | Means |
|---|---|
| $R_i$ | Total data packets received by $UAV_i$ |
| $S_i$ | Total data packets sent by $UAV_i$ |
| N | Number of scenarios executed |

**Detection Rate:** The definition of this measure is as follows: the ratio of intruder things to total intruder things that have been correctly identified as intruder things [23]. This measure has been analyzed using Eq. (10). Also, the meaning of the variables used in Eq. (10) is explained in Table 2.

$$ADR = \left(\frac{ATP}{ATP + AFN}\right) * 100$$

$$Where: \quad All = ATP + ATN + AFP + AFN \tag{10}$$

$$AFP = \left(\frac{AFP}{AFP + ATN}\right); \qquad AFN = \left(\frac{ATP + ATN}{All}\right)$$

$$ATP = \left(\frac{ATP}{ATP + AFN}\right); \qquad ATN = \left(\frac{ATN}{ATN + AFP}\right)$$

**Table 2** Definition of measures.

| Measure | Definition |
|---|---|
| AFP | The number of normal nodes falsely detected as intruder nodes [29]. |
| AFN | The number of intruder nodes detected as the normal nodes divided by the total number of intruder nodes. |
| ATN | The number of normal nodes correctly detected as the normal nodes divided by the total number of normal nodes. |
| ATP | The number of normal nodes falsely detected as the intruder nodes divided by the total number of normal nodes [30]. |

## 4.2 Simulation results and analysis

In this subsection, simulation results are shown for all measures. The proposed DFA-RPL method has been simulated and its performance evaluated in NS-3 Simulator on Linux Ubuntu 14 LTS. Because the data extracted from the simulation is correct and logical, the most important parameters of things such as MAC Layer, etc. are used. The rest of the parameters used in the simulator are listed in Table 3.

Details of the parameters used in the three scenarios are given in Table 4. The only difference between the three scenarios is the intruder thing rate. The rest of the parameters are considered the same.

**Table 3** Details of the parameters used in the simulation

| Parameters | Value |
|---|---|
| Number of nodes | 500 |
| Packet size | 512 bytes |
| Transmission range | 20 M |
| Transport | UDP/IPv6 |
| Intruder thing ratio | 10%, 15%, 20% |
| Channel type | Wireless |
| MAC | MAC/802.11 |
| Traffic type | Constant Bit Rate |
| Transmission layer | UDP (User Datagram Protocol) |
| Time | 2000 |

**Table 4** Important parameters of all three scenarios

| Scenario #1 | | Scenario #2 | |
|---|---|---|---|
| Number of nodes | 500 | Number of nodes | 500 |
| Intruder rate (%) | 10 | Intruder rate (%) | 15 |
| Topology (m x m) | 3000 x 3000 | Topology (m x m) | 3000 x 3000 |
| Time | 2000 | Time | 2000 |
| Scenario #3 | | | |
| Number of nodes | 500 | | |
| Intruder rate (%) | 20 | | |
| Topology (m x m) | 3000 x 3000 | | |
| Time | 2000 | | |

**Packet Delivery Rate:** The simulation results of all three methods DFA-RPL, REATO, and IRAD under PDR criteria are shown in Table 5 and Figure 3. By increasing the value of the Number of nodes variable from 50 to 500, the PDR criterion of all three methods has also increased. The DFA-RPL method is significantly better than the REATO and IRAD methods. The reason for improving the proposed DFA-RPL method is that in the proposed method, it first builds a cluster and selects the most appropriate node as a cluster head in DODAG, then, due to the vulnerability of the RPL protocol to Flooding attacks, the ant colony optimization (ACO) to detect attacks Use Flooding to quickly identify intrusive nodes and exclude routing operations. Therefore, as can be seen from Table 5 and Figure 3, the proposed method is approximately 29% superior to the REATO and IRAD methods in the PDR criterion.

**Table 5** Packet Delivery Rate all 3 methods against the number of nodes.

| Misbehaving Things ratio (%) | Packet Delivery Rate (%) | | |
|---|---|---|---|
| | IRAD | REATO | DFA-RPL |
| 0 | 84.4 | 86.4 | 98.4 |
| 0.05 | 76.1 | 80.1 | 95.4 |
| 0.1 | 70.1 | 73.1 | 93.1 |
| 0.15 | 62.3 | 68.3 | 92.1 |
| 0.2 | 55.13 | 60.13 | 89 |
| 0.25 | 49.2 | 53.2 | 88 |
| 0.3 | 35.23 | 45.23 | 86 |

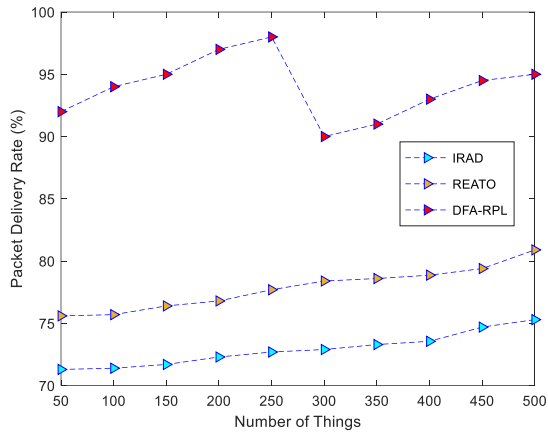

A. Number of nodes (10% intruder)

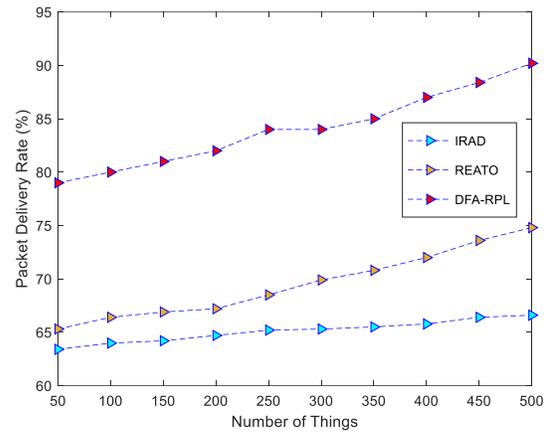

B. Number of nodes (15% intruder)

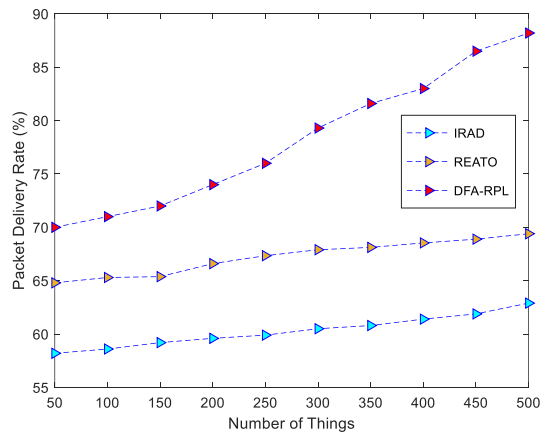

C. Number of nodes (20% intruder)

**Fig. 3** Comparison of the DFA-RPL proposed scheme, REATO and IRAD approaches in term of Packet Delivery Rate.

**Detection Rate:** Table 6 and Figure 4 show the simulation results of all three DFA-RPL, REATO and IRAD methods under the Detection Rate criterion. By increasing the value of the Number of nodes variable from 50 to 500, the PDR criterion of all three methods has also increased. The DFA-RPL method is significantly better than the REATO and IRAD methods. The reason for the improvement of the proposed DFA-RPL method is that the proposed method in the first part has done clustering and selecting the cluster to create DODAG and in the second part the Flooding attack detection has been done using ACO algorithm. This prevents intrusive nodes from working and quickly stops them from working on the network. Therefore, as can be seen from Table 6 and Figure 4, the proposed method is approximately 24% superior to the REATO and IRAD methods in the PDR criterion.

**Table 6** Detection Rate all 3 methods against the number of nodes.

|  | Detection Rate (%) | | |
| --- | --- | --- | --- |
| Misbehaving Things ratio (%) | IRAD | REATO | DFA-RPL |
| 0 | 91.63 | 90.2 | 95 |
| 0.05 | 89.49 | 88.57 | 94 |
| 0.1 | 80.46 | 81.8 | 93 |
| 0.15 | 73.35 | 76.37 | 92 |
| 0.2 | 63.19 | 70.43 | 91 |
| 0.25 | 50.34 | 66.16 | 89 |
| 0.3 | 46.14 | 60.67 | 80 |

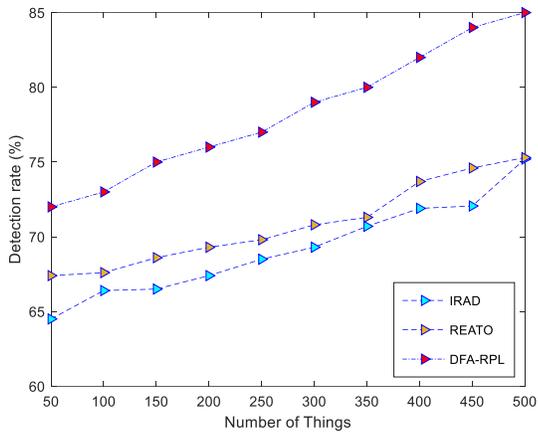

A. Number of nodes (10% intruder)

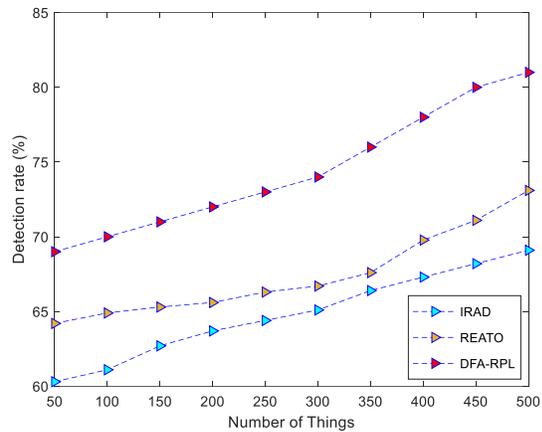

B. Number of nodes (15% intruder)

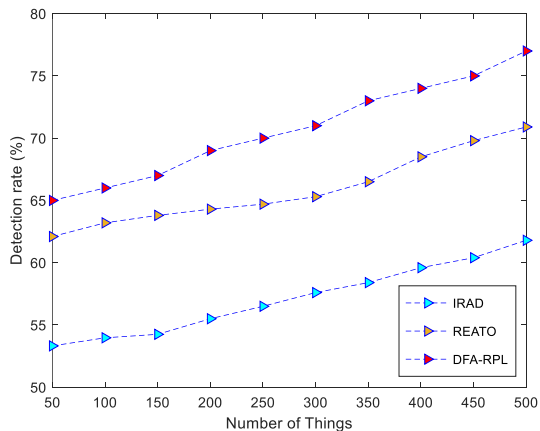

C. Number of nodes (20% intruder)

**Fig. 4** Comparison of the DFA-RPL proposed scheme, REATO and IRAD approaches in term of Detection Rate.

**False Positive Rate:** This criterion indicates that the method in question mistakenly identified a normal node as a malicious node. The simulation results are shown in Table 7 and Figure 5. According to the diagrams, it is clear that the proposed method in the False Positive Rate criterion has a much better performance than the other two methods. The reason for the superiority of the proposed method is that in the first part, it is used to create an optimal DODAG that has the best nodes as a parent. The task of the second part is to detect malicious objects, which are most likely Flooding attack nodes, which try to disrupt network operations by sending a large number of messages. These malicious nodes are detected using the ACO algorithm and using five steps. And its malicious activity in the network is prevented by saying that if the node is found to be malicious, that node will be placed on the detention list and will be quarantined for a certain period of time. The proposed DFA-RPL scheme can efficiently create an optimal DODAG and manage the Flooding attack on the network.

**Table 7** False Positive Rate all 3 methods against the number of nodes.

| Misbehaving Things ratio (%) | False Positive Rate (%) | | |
|---|---|---|---|
| | IRAD | REATO | DFA-RPL |
| 0 | 8.3 | 9.25 | 3.6 |
| 0.05 | 10.31 | 12.24 | 4.8 |
| 0.1 | 19.05 | 19.01 | 6.3 |
| 0.15 | 27 | 23.35 | 7.9 |
| 0.2 | 34.38 | 28.62 | 8.3 |
| 0.25 | 47.6 | 31.88 | 9.15 |
| 0.3 | 52.67 | 39.09 | 10.5 |

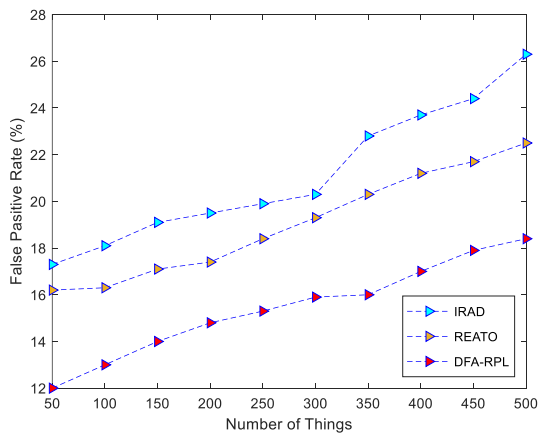

A. Number of nodes (10% intruder)

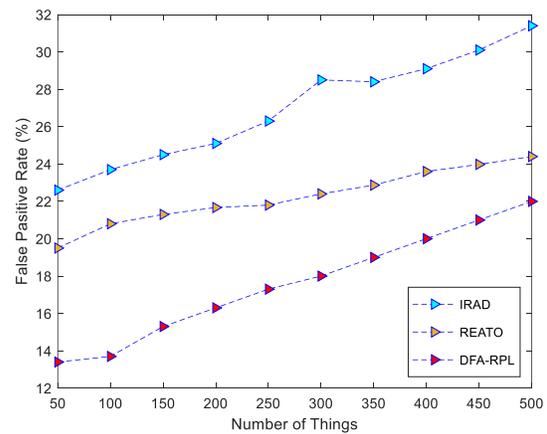

B. Number of nodes (15% intruder)

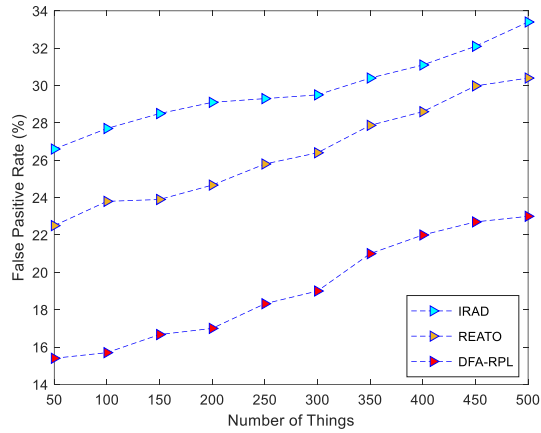

C. Number of nodes (20% intruder)

**Fig. 5** Comparison of the DFA-RPL proposed scheme, REATO and IRAD approaches in term of False Positive Rate.

**False Negative Rate:** Table 8 and Figure 6 show the simulation results of all three DFA-RPL, REATO and IRAD methods under the False Negative Rate criterion. By increasing the value of the Number of nodes variable from 50 to 500, the False Negative Rate criterion of all three methods has increased, but the increase of the proposed method is much lower than the other two methods. Our proposed method is more efficient than the REATO and IRAD methods under flooding attack, because it not only detects the malicious node in advance, but also isolates the node in the network and re-nodes the accused node after the penalty period has elapsed. Enters the operation cycle. DFA-RPL increases the throughput of the entire network. The main advantage of using DFA-RPL is that after a reasonable penalty, the accused node can be considered as a normal node again in the network.

**Table 8** False Negative Rate all 3 methods against the number of nodes.

|  | False Negative Rate (%) | | |
| --- | --- | --- | --- |
| Misbehaving Things ratio (%) | IRAD | REATO | DFA-RPL |
| 0 | 7.93 | 9.005 | 3.6 |
| 0.05 | 9.4 | 10.08 | 4.5 |
| 0.1 | 12.4 | 11.3 | 4.9 |
| 0.15 | 14 | 13.37 | 5.3 |
| 0.2 | 17 | 14.6 | 5.9 |
| 0.25 | 19 | 15.7 | 6.9 |
| 0.3 | 23 | 17.6 | 8.2 |

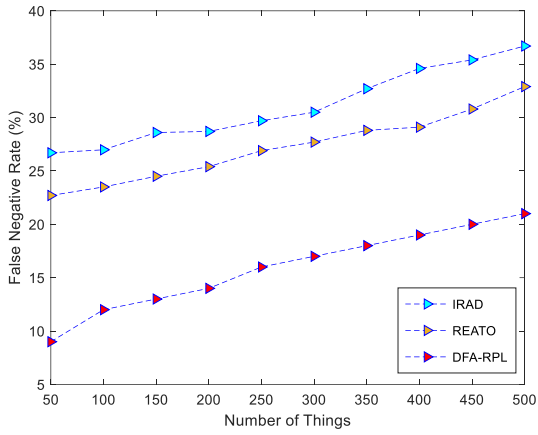

A. Number of nodes (10% intruder)

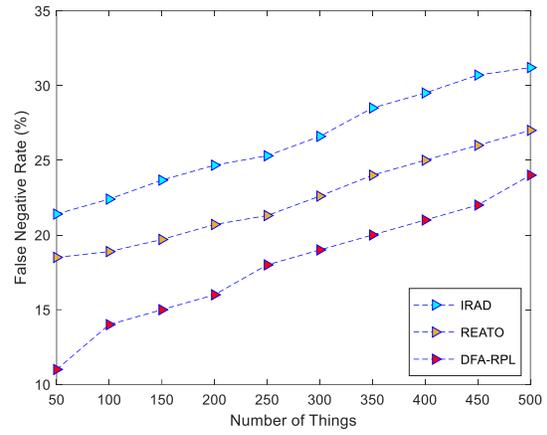

B. Number of nodes (15% intruder)

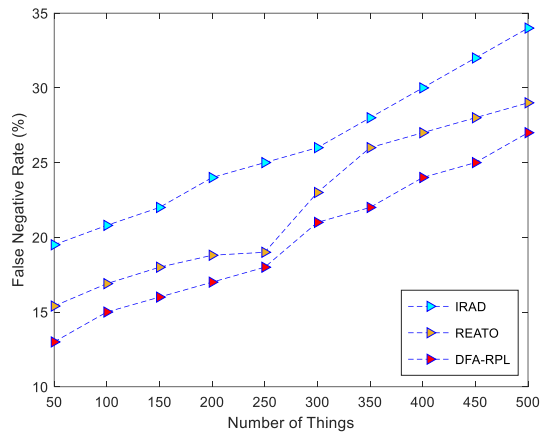

C. Number of nodes (20% intruder)

**Fig. 6** Comparison of the DFA-RPL proposed scheme, REATO and IRAD approaches in term of False Negative Rate.

## 5 Conclusion

The proposed DFA-RPL method consists of two main parts. In the first part, clustering is performed, and the cluster head is selected to create the DODAG. In the second part, the flooding attack is detected using the ACO algorithm. The first part is used to create an optimal DODAG with the best nodes as parents. The second part is responsible for detecting malicious objects that are most probably malicious nodes in a flooding attack and attempt to disrupt the network operation by transmitting a large number of attacks. These malicious nodes are detected using the ACO algorithm and within 5 steps, and their malicious activity is prevented. This is done as follows: if the node is recognized as malicious, it is placed inside the detention list and quarantined for a specific period. The proposed DFA-RPL scheme can efficiently create an optimal DODAG and manage the flooding attack in the network. To demonstrate the effectiveness of the proposed method (DFA-RPL), we compared this method with the DSR protocol under flooding attack using the NS-2 simulator and used the criteria of number of lost packages, package arrival rate, end-to-end delay, and throughput to show the performance of our method. Our proposed method is more efficient compared to the REATO and

IRAD methods under flooding attack since not only does it detect the malicious node beforehand but also isolates the node in the network and monitors the suspect node again after the end of the penalty time. DFA-RPL increases the throughput of the whole network. The main advantage of using DFA-RPL is that after a reasonable penalty, the suspect node can be considered a normal node again in the network.

**Declaration:**

### Funding:

No funds, grants, or other supports was received.

### Competing interests:

The authors declare that they have no competing interests.

### Availability of data and materials:

Not applicable

### Code availability:

Not applicable

### Ethical approval:

Not applicable

### Consent to participate:

Not applicable

### Consent to publish:

Not applicable

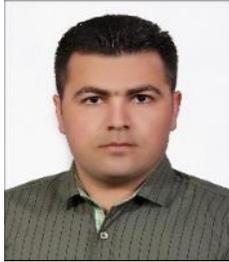

**Reza Fotohi** is currently a Ph.D. candidate in the Faculty of Computer Science and Engineering, Shahid Beheshti University, Tehran, IRAN. His research interests include the IoT Security, AI Security, Network, Cryptography, Blockchain, Deep Learning. He has over 8 years' research experience in computer networks and mobile/wireless networks. He was also the recipient and co-recipient of various awards for research publications. He is the author and co-author of more than 20 technical journal and conference papers. He is also a member and reviewer in many journals and conferences such as IEEE Communications Surveys and Tutorials, IEEE Access, Applied Soft Computing, Computer Communications, Artificial Intelligence Review, Iran Journal of Computer Science, The Journal of Supercomputing, Neural Processing Letters, Human-centric Computing, and Information Sciences, Wireless Personal Communications, Journal of Ambient Intelligence and Humanized Computing, Journal of Grid Computing, National Academy Science Letters, Journal of Intelligent & Robotic Systems, International Journal of Communication Systems, Transactions on Emerging Telecommunications Technologies, IET Signal Processing, Electronics Letters, KSII Transactions on Internet and Information Systems, The Turkish Journal of Electrical Engineering & Computer Sciences, Iranian Journal of Fuzzy Systems (IJFS), and Jordanian Journal of Computers and Information Technology. His papers have more than 736 citations with 19 h-index and 23 i10-index.

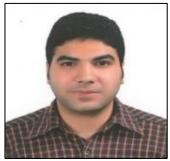

**Hossein Pakdel** received his B.S. degree in software engineering from Islamic Azad University, Isfahan (Khorasgan) branch in 2012. He received the M.S. degree in software engineering from Islamic Azad University, central Tehran branch in 2016. His current research focuses on energy conservation in linear wireless sensor networks. His research interests include wireless sensor networks, distributed systems, and software engineering.